\documentclass[conference]{IEEEtran}
\IEEEoverridecommandlockouts
\usepackage{cite}
\usepackage{adjustbox}
\usepackage[ruled,vlined]{algorithm2e}
\usepackage{amsmath,amssymb,amsfonts}
\usepackage{algorithmic}
\usepackage{graphicx}
\usepackage{textcomp}
\usepackage[colorlinks=false, pdfborder={0 0 0}, breaklinks]{hyperref}
\usepackage{xcolor}
\usepackage{balance}
\usepackage{enumitem}
\usepackage{setspace}
\usepackage{wrapfig}
\usepackage{listings}
\usepackage{color}
\usepackage{subcaption}
\usepackage{chemmacros}
\usepackage[longtable]{multirow}
\newcommand{\rf}[1]{\textcolor{black}{#1}}
\newcommand{\sam}[1]{\textcolor{black}{#1}}
\usepackage{pifont}
\usepackage{xcolor}
\usepackage{tikz}
\usepackage{verbatim} 
\usepackage[accsupp]{axessibility}

\def\BibTeX{{\rm B\kern-.05em{\sc i\kern-.025em b}\kern-.08em
    T\kern-.1667em\lower.7ex\hbox{E}\kern-.125emX}}

\usepackage{amsmath}

\DeclareMathOperator*{\argmin}{arg\,min}
\usepackage{mathtools}

\usepackage{tikz}
\usetikzlibrary{shapes.geometric, arrows}
\usetikzlibrary{fit}
\usetikzlibrary{calc}
\usetikzlibrary{shapes.geometric, arrows}
\tikzstyle{arrow} = [thick,->,>=stealth]

\usepackage{ctable}
\newcommand{\ie}{\emph{i.e.}, }
\newcommand{\eg}{\emph{e.g.}, }

\newcommand{\cf}{\emph{cf.}\xspace}
\newcommand{\HAS}{\emph{HTTP Adaptive Streaming }}
\newcommand{\DASH}{\emph{Dynamic Adaptive Streaming over HTTP }}
\newcommand{\HEVC}{\emph{High Efficiency Video Coding}\xspace}
\newcommand{\AVC}{\emph{Advanced Video Coding}\xspace}
\newcommand{\AVI}{\emph{Alliance for Open Media Video 1}\xspace}

\newcommand{\jnd}{\emph{Just Noticeable Difference}\xspace}

\newcommand{\mcbe}{\texttt{MCBE}\xspace}
\newcommand{\etpsplus}{\texttt{JTPS}\xspace}
\newcommand{\opte}{\texttt{OPTE}\xspace}

\newcommand{\EY}{$E_{\text{Y}}$}
\newcommand{\LY}{$L_{\text{Y}}$}
\newcommand{\h}{$h$}

\usepackage{lipsum} 
\usepackage{tikz,xcolor,hyperref}
\definecolor{lime}{HTML}{A6CE39}
\DeclareRobustCommand{\orcidicon}{%
	\begin{tikzpicture}
	\draw[lime, fill=lime] (0,0) 
	circle [radius=0.16] 
	node[white] {{\fontfamily{qag}\selectfont \tiny ID}};
	\draw[white, fill=white] (-0.0625,0.095) 
	circle [radius=0.007];
	\end{tikzpicture}
	\hspace{-2mm}
}

\foreach \x in {A, ..., Z}{%
	\expandafter\xdef\csname orcid\x\endcsname{\noexpand\href{https://orcid.org/\csname orcidauthor\x\endcsname}{\noexpand\orcidicon}}
}

\begin{document}
\bstctlcite{IEEEexample:BSTcontrol}

\title{Energy-Efficient Multi-Codec Bitrate-Ladder Estimation for Adaptive Video Streaming}

%\author{Anonymous VCIP submission}
%\iffalse
\author{\IEEEauthorblockN{Vignesh V Menon\IEEEauthorrefmark{1}\IEEEauthorrefmark{2}, Reza Farahani\IEEEauthorrefmark{2}, Prajit T Rajendran\IEEEauthorrefmark{3},\\ Samira Afzal\IEEEauthorrefmark{2}, Klaus Schoeffmann\IEEEauthorrefmark{2}, Christian Timmerer\IEEEauthorrefmark{2}
}

\IEEEauthorblockA{\IEEEauthorrefmark{1} Video Communication and Applications Department, Fraunhofer HHI, Berlin, Germany}
\IEEEauthorblockA{\IEEEauthorrefmark{2} Alpen-Adria-Universit{\"a}t, Klagenfurt, Austria}
\IEEEauthorblockA{\IEEEauthorrefmark{3} Universite Paris-Saclay, CEA, List, F-91120, Palaiseau, France}
}
%\fi
\maketitle
\begin{abstract}
%\rf{The popularity of streaming segmented videos through the Hypertext Transfer Protocol (HTTP) is on the rise. \HAS (HAS)-based solutions serve client requests segment by segment in various representations. A set of representations, known as the \textit{bitrate ladder}, expresses this choice as an ordered list of bitrate-resolution pairs.} 
\rf{With the emergence of multiple modern video codecs, streaming service providers are forced to encode, store, and transmit bitrate ladders of multiple codecs separately, consequently suffering from additional energy costs for encoding, storage, and transmission.} 
\rf{To tackle this issue,} we introduce an online energy-efficient \underline{M}ulti-\underline{C}odec \underline{B}itrate ladder \underline{E}stimation scheme (\mcbe) for adaptive video streaming applications. In \mcbe, quality representations within the bitrate ladder of new-generation codecs \sam{(\eg \HEVC (HEVC), \AVI (AV1))} that lie below the predicted rate-distortion curve of the \AVC~(AVC) codec are removed. Moreover, perceptual redundancy between representations of the bitrate ladders of the considered codecs is also minimized \rf{based on a} \rf{\textit{Just Noticeable Difference} (JND)} threshold. Therefore, random forest-based models predict the VMAF score of bitrate ladder representations of each codec. In a live streaming session where all clients support the decoding of AVC, HEVC, and AV1, \mcbe achieves impressive results, reducing cumulative encoding energy by 56.45\%, storage energy usage by 94.99\%, and transmission energy usage by 77.61\% (considering a JND of six VMAF points). These energy reductions are in comparison to a baseline bitrate ladder encoding based on current industry practice.
\end{abstract}
\begin{IEEEkeywords}
\rf{HTTP Adaptive Streaming; Multi-Codec~Streaming; Per-Title~Encoding; Energy-Aware~Streaming; Just~Noticeable~Difference}.
\end{IEEEkeywords}

\section{Introduction}
\begin{figure}[t]
\centering
\includegraphics[width=.47\textwidth]{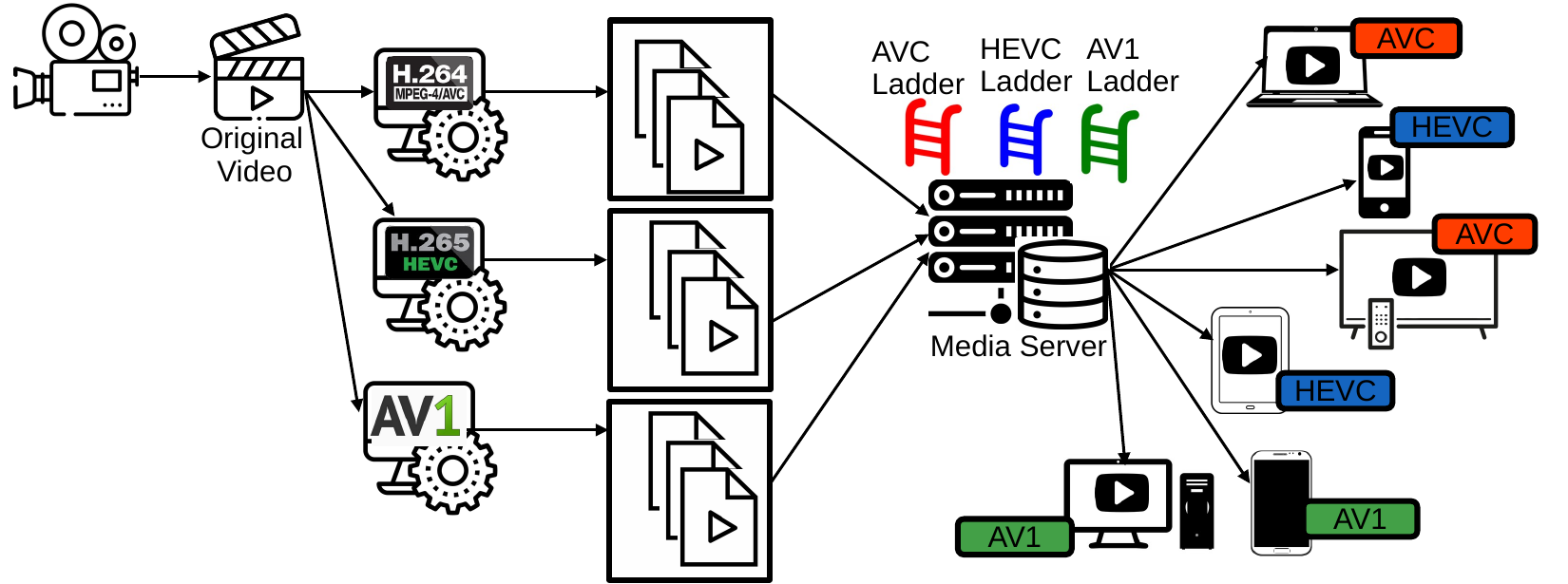}
%\vspace{0.6em}
\caption{\rf{An example of a multi-codec streaming system.}}
\label{fig:mc_intro}
\vspace{-0.75em}
\end{figure}

The emergence of novel video formats and standards has facilitated content delivery across various platforms and devices. \HAS (HAS) delivery systems, such as those based on the MPEG \DASH (DASH)~\cite{DASH_IEEE} standard or Apple \textit{HTTP Live Streaming} (HLS)~\cite{HLS}, have emerged as the dominant technologies utilized by service providers to deliver live video content~\cite{farahani2022hybrid,farahani2022ararat}. \rf{In such systems}, each codec requires its own set of \rf{representations}, \ie bitrate ladders~\cite{DASH_Survey, multi_codec_ct_ref}. For example, \emph{Advanced Video Coding} (AVC)~\cite{AVC} and \emph{High Efficiency Video Coding} (HEVC)~\cite{HEVC} have distinct bitrate ladders. Initially, streaming services used AVC for wider device compatibility~\cite{yuriy_multicodec_ref}. However, as newer devices with HEVC and \emph{Alliance for Open Media Video 1} (AV1)~\cite{av1_ref} support becomes prevalent, HEVC and AV1-encoded bitrate ladder representations are introduced. Recent years have developed new formats such as \emph{Versatile Video Coding} (VVC)~\cite{vvc_ref}, \emph{Essential Video Coding} (EVC)~\cite{evc_ref}, and \emph{Low Complexity Enhancement Video Coding} (LCEVC)~\cite{lcevc_ref}. Over time, streaming systems have evolved to accommodate multiple codecs, with older devices relying solely on AVC, some newer devices using HEVC streams, and certain devices supporting both AVC and HEVC, including seamlessly switching between them~\cite{yuriy_multicodec_ref} (\cf Fig.~\ref{fig:mc_intro}). Handling such multi-codec deployments requires generating ABR bitrate ladders of each codec separately, considering the range of codecs to be supported by the receiving device population based on their decoding capabilities~\cite{reznik2019optimal}.
%%%
%%%%
\begin{figure}[t]
\centering
\begin{subfigure}{0.241\textwidth}
\centering
\includegraphics[width=\linewidth]{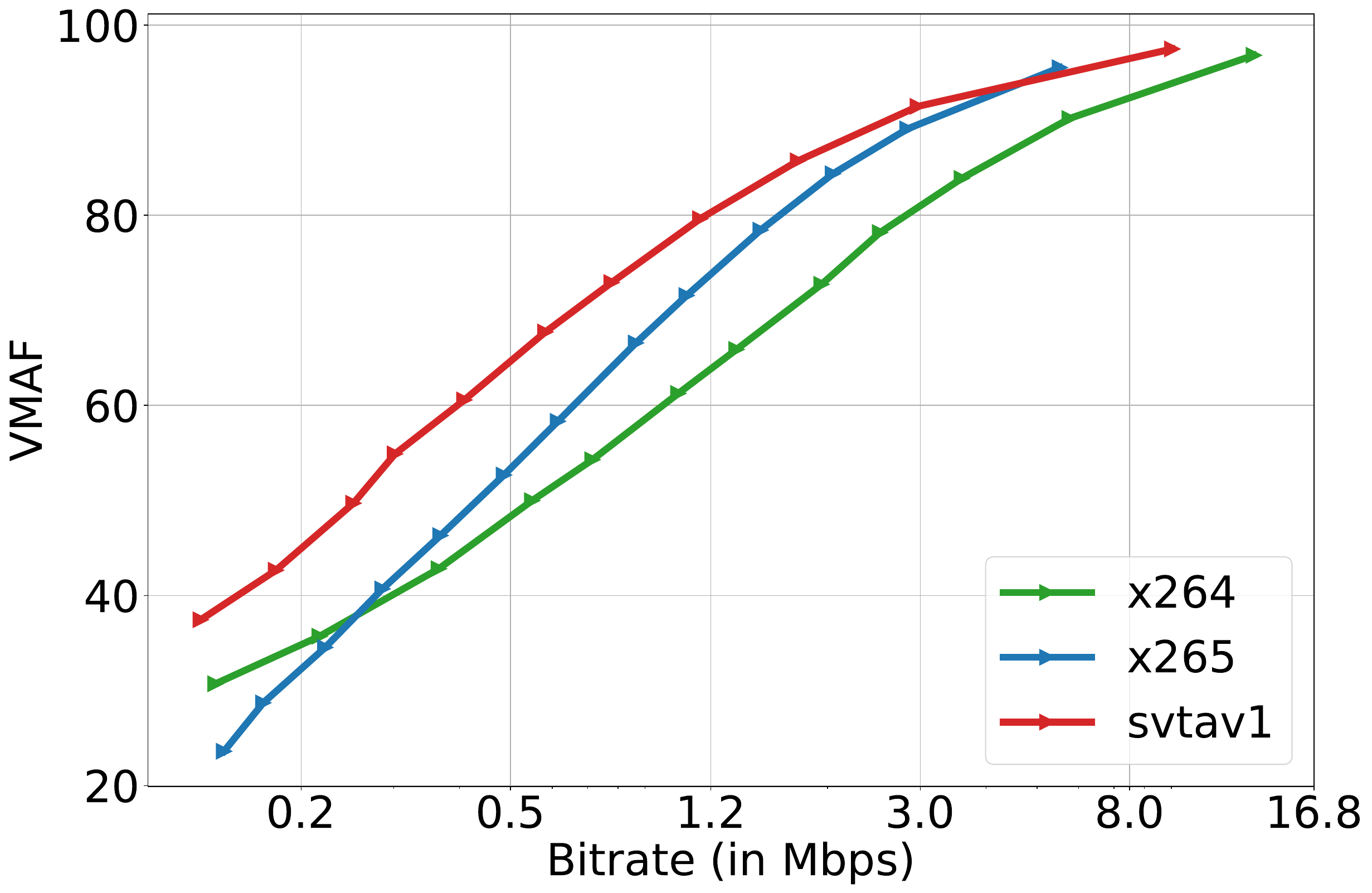}
\caption{\textit{Basketball\_s000}}
\end{subfigure}
\hfill
\begin{subfigure}{0.241\textwidth}
\centering
\includegraphics[width=\linewidth]{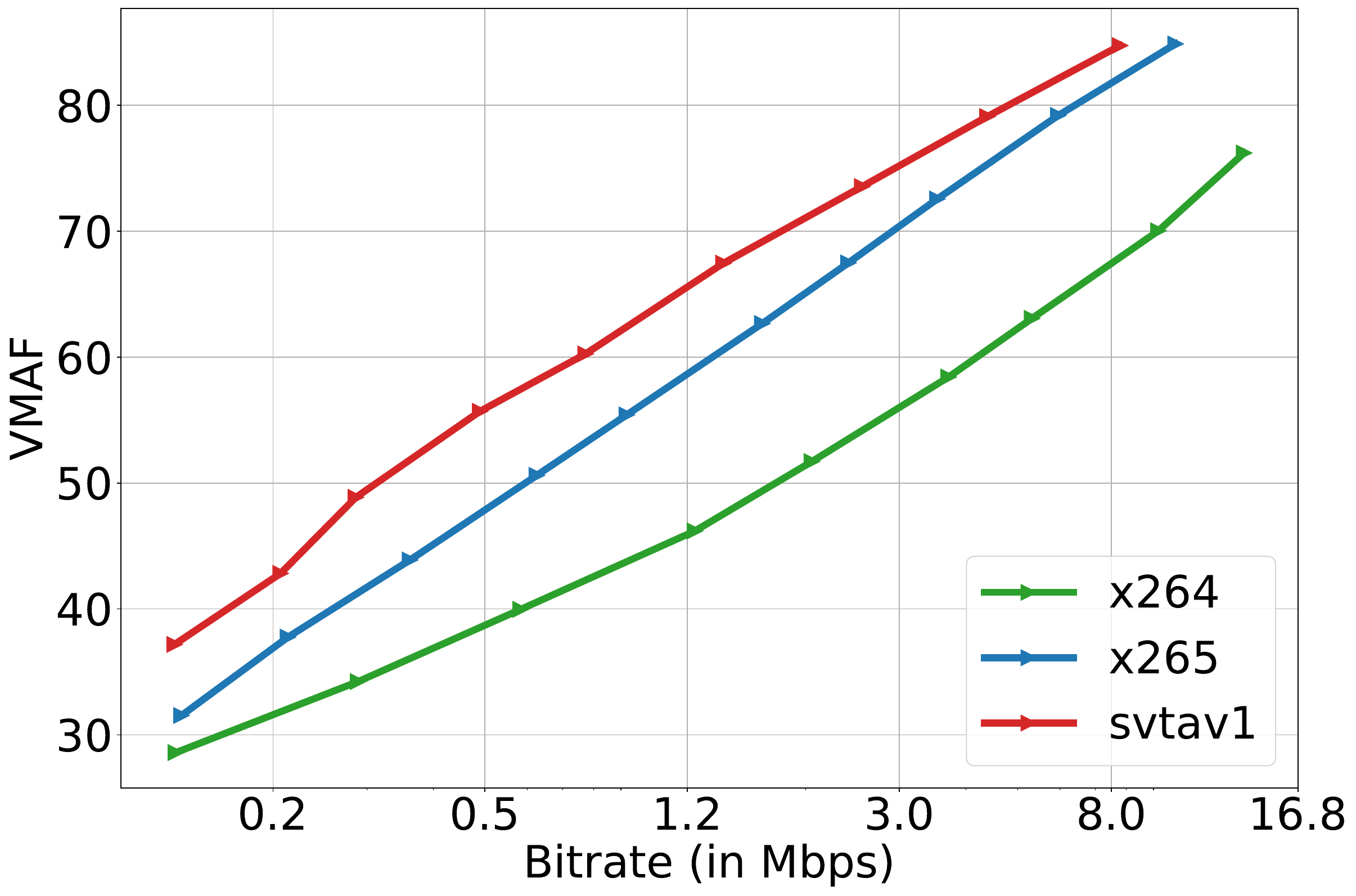}
\caption{\textit{Riverbank\_s000}}
\end{subfigure}
%\vspace{-1.05em}
\caption{Rate-distortion (RD) curves of representative sequences of VCD dataset~\cite{VCD_ref}, encoded with \etpsplus bitrate ladder~\cite{jtps_ref} for x264~\cite{x264_ref}, x265~\cite{x265_ref}, and svtav1~\cite{svtav1_ref} encoders.}
\vspace{-1.5em}
\label{fig:vision}
\end{figure}
%%%
\begin{figure*}[t]
    \centering
    \includegraphics[width=.77\linewidth]{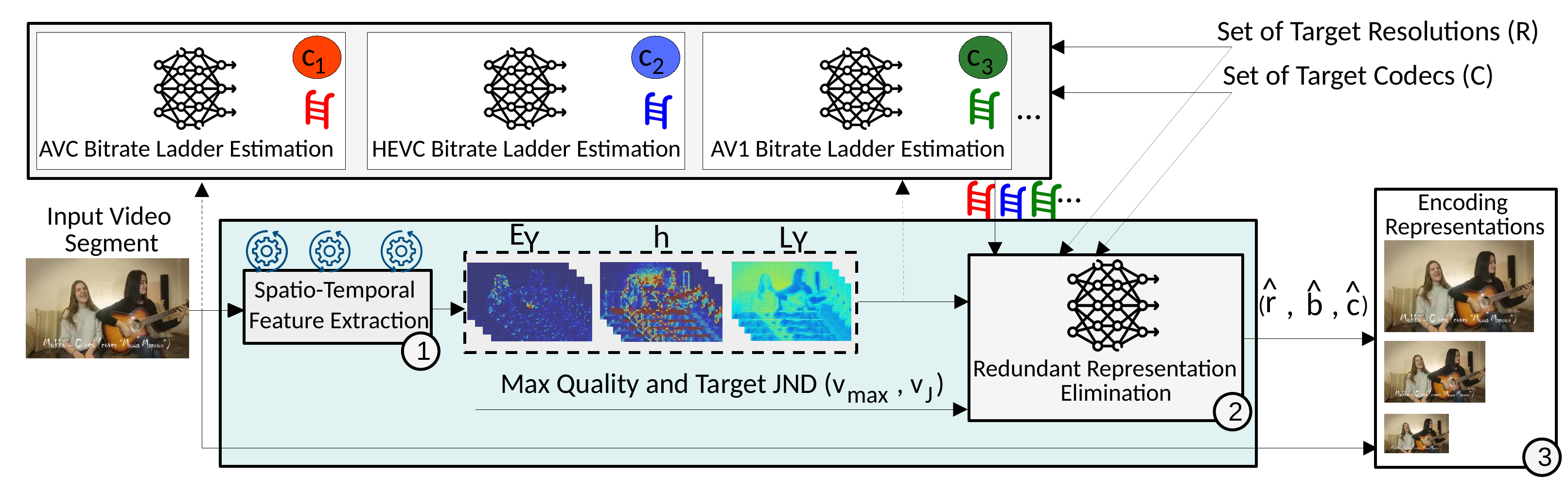}
    \vspace{-.2cm}
    \caption{Online encoding using \mcbe envisioned in this paper for adaptive video streaming.}
    \vspace{-1.1em}
    \label{fig:mcps_scheme}
\end{figure*}
%%%

The estimation of a multi-codec bitrate ladder, as proposed in this paper, is based on the fact that, in some cases, the compression efficiency of AVC is better than new-generation video codecs, \ie at low bitrates~\cite{codec_comp_ref, hevc_av1_comp1}. Furthermore, the compression efficiency of codecs saturates at very high target bitrates, as they become similar to lossless coding. The regions where each codec performs better than others depend on the \textit{complexity of the video content}~\cite{vca_ref}. An example is shown in Fig.~\ref{fig:vision} where the cross-over bitrate between the quality of x264~\cite{x264_ref} and x265~\cite{x265_ref} is at approximately $b_{1} =0.3$\,Mbps for \textit{Basketball\_s000}, while the cross-over bitrate between the quality of x265 and svtav1 is at approximately $b_{2} = 5.0$\,Mbps. This means that, at bitrates lower than $b_{1}$, x264 outperforms x265, while at bitrates higher than $b_{2}$, x265 outperforms svtav1. On the other hand, for \textit{Riverbank\_s000}, svtav1 remains superior throughout the bitrate range. This is because old-generation codecs may excel in scenarios where the content does not leverage the advanced coding tools and techniques introduced by the new-generation codecs. x265 encoding of \textit{Basketball\_s000} at bitrates lower than $b_{1}$ can be eliminated, as clients can be served with x264 representations (due to comprehensive support of AVC).% This is because AVC is supported by all clients to-date. %An optimized multi-codec ladder also offers \textit{(i)} simplified management, \textit{(ii)} improved interoperability, \textit{(iii)} greater flexibility and adaptability in delivering content to diverse network conditions and client devices, \textit{(iv)} overcome compatibility issues that may arise when using indepedent ladders of multiple codecs~\cite{reznik2019optimal}. Another advantage of an optimized multi-codec bitrate ladder is the elimination of the need to transcode or re-encode content between different codecs, which might lead to quality degradation and increased processing requirements.

Encoding video content into multiple representations in various bitrate-resolution pairs for each codec results in substantial computational workload and energy consumption~\cite{koziri2018efficient}. Additionally, the storage and transmission of these representations further contribute to the overall energy consumption~\cite{baliga2010green}. When unnecessary high-bitrate representations (of new-generation codecs) are eliminated, the energy consumption of the streaming system is significantly reduced~\cite{daniele_ds_ref}. This is because the energy consumption of AVC is significantly lower than that of new-generation video codecs~\cite{codec_comp_ref1,hevc_av1_comp}. As video streaming continues to grow in popularity and usage, finding energy-efficient solutions to optimize the multi-codec bitrate ladder becomes crucial to mitigate the environmental impact and reduce operational costs for streaming service providers~\cite{ds_paper_ref}.

In this light, this paper proposes an online \underline{M}ulti-\underline{C}odec \underline{B}itrate ladder \underline{E}stimation (\mcbe) scheme for adaptive video streaming applications. A lightweight algorithm is proposed to \emph{eliminate redundant representations of the bitrate ladders of new-generation video codecs}, based on their predicted perceptual quality. Therefore, random forest models~\cite{rf_ref} are trained to estimate the VMAF score of each representation based on low-complexity spatio-temporal features of the input video segment.  Note that other quality metrics can be envisioned, which are subject to our future work. 
%algorithm to \emph{eliminate redundant representations of the bitrate ladders of new-generation video codecs} is proposed, based on \DCT (DCT)-energy-based low-complexity spatio-temporal features of the video segment. 
\rf{When} AVC performs better (in terms of perceptual quality) than or is \rf{identical} as HEVC and/or \rf{another} new-generation codec (\eg AV1) in a bitrate range, the corresponding new-generation codec representations are eliminated from the bitrate ladder\rf{. This is because} clients can be served with AVC representations with better RD performance. Moreover, the bitrate ladder representations with a perceptual quality difference within a given \jnd (JND)~\cite{jnd_ref} threshold are eliminated. Finally, it is worth noting that \mcbe can be used in conjunction with state-of-the-art bitrate-ladder prediction schemes~\cite{ml_bl_pred, gnostic, opte_ref, jtps_ref}). 
%It is demonstrated that, using \mcbe reduces the energy consumption of encoding, storage and transmission by up to 54.34\%, 95.30\%, and 78.32\%, respectively, compared to independent encodings of x264, x265 and svtav1 using state-of-the-art \opte bitrate ladder prediction~\cite{opte_ref}. When \etpsplus~\cite{jtps_ref} bitrate ladder prediction is used, reductions in energy consumption are 27.80\%, 40.12\%, and 22.62\%, respectively.

\section{\mcbe architecture}
\label{sec:mcps_framework}
\sam{Adaptive video streaming systems often use bitrate-ladder prediction methods to enhance the Quality of Experience (QoE) for users~\cite{netflix_paper, gnostic, faust_ref}}. The architecture of the proposed \mcbe scheme is shown in Fig.~\ref{fig:mcps_scheme}. \mcbe receives input bitrate ladders for each codec, \eg $c_1$, $c_2$, and $c_3$ for AVC, HEVC, and AV1 codecs, respectively. Other codecs may be envisioned as part of future work but are supported by the current architecture. It extracts DCT-energy-based features and eliminates redundant representations based on the predicted quality metric (\ie VMAF in this paper) of each representation. 
\mcbe comprises three phases (\cf Fig.~\ref{fig:mcps_scheme}): 
 \newcommand*\circled[1]{\tikz[baseline=(char.base)]{\node[shape=circle,draw,inner sep=1pt] (char) {#1};}}
 \begin{enumerate}[label=\protect\circled{\color{black}\arabic*}, topsep=1pt, leftmargin=*]
  \item Spatio-temporal feature extraction (Section~\ref{sec:features})
  \item Redundant representation elimination (Section~\ref{sec:rep_elim})
  \item Encoding of the segments using the selected bitrate-resolution pairs of each codec
\end{enumerate}

%\mcbe first extracts the spatio-temporal features of the input video segment. The features, along with $b_{min}$, $b_{max}$, $v_{max}$, $R$ are input to the considered bitrate ladder estimation scheme for each codec. 
%VMAF is predicted for each representation selected by the bitrate ladder prediction scheme. Perceptually-redundant representations, \ie representations which yield VMAF difference of $v_{J}$ are eliminated. Furthermore, representations of new-generation codecs which are below the rate-distortion curve of AVC encoding are also eliminated.  
%This ensures that the scheme can be tuned to accommodate the optimized number of representations in the bitrate ladder by considering the maximum bitrate and VMAF supported in the bitrate ladder of the streaming service provider. $R$ is input to the scheme to ensure that only the resolutions supported by the streaming service provider are selected to generate the encoding set.
%    takes  and predicts its corresponding VMAF. The VMAF scores for the remaining representations are determined by increasing VMAF (until  or  is reached) of the previous point of the bitrate ladder by one JND. These VMAF values are then used to predict their corresponding bitrate-resolution pairs.
%Finally, the bitrate ladder consists of only the determined codec-resolution-bitrate combinations, eliminating the need to encode and store all representations of all considered codecs. 

\subsection{Spatio-Temporal Feature Extraction}
\label{sec:features}
\mcbe uses the following DCT-energy-based features~\cite{dct_ref}, extracted using open-source VCA v2.0 video complexity analyzer~\cite{vca_ref} for every segment: 
\begin{enumerate}
\item Average luma texture energy (\EY)
\item Average gradient of the luma texture energy (\h)
\item Average luminescence (\LY)
\end{enumerate}

\subsection{Redundant Representation Elimination} 
\label{sec:rep_elim}
In this paper, \rf{the} VMAF \rf{score} $v_{r_{t}, b_{t}, c}$ of the $t^{th}$ representation of the codec $c$ is modeled as a function of the video content complexity features and the target representation (\ie resolution $r_{t}$ and bitrate $b_{t}$)~\cite{tqpm_ref,vqa_vca_ref}, as shown in the following equation: 
\begin{equation}
v_{r_{t}, b_{t}, \rf{c}} = f_{\text{V}}(E_{\text{Y}}, h, L_{\text{Y}}, r_{t}, b_{t}, \rf{c}) 
\end{equation}
Random forest models~\cite{rf_ref} which are hyperparameter-tuned with the parameters \textit{min\_samples\_leaf}=1, \textit{min\_samples\_split}=2, \textit{n\_estimators}= 100, and \textit{max\_depth}=14 are trained for each codec $c \in \mathcal{C}$ and resolution $r \in \mathcal{R}$ to predict VMAF. Input to the model for each codec-resolution are [\EY, \h, \LY, $b$].
\setlength{\textfloatsep}{2pt}
\begin{algorithm}[t]
{%\footnotesize
\small
\textbf{Inputs:}\\
$M$~: number of supported codecs\\
$\mathcal{C}$~: set of all codecs {$c_{1}$, $c_{2}$...$c_{M}$} in order of priority\\
$N_c$~: number of representations for codec $c$ \\
$(\hat{r}_t,\hat{b}_t,c)$ pairs $\forall c \in \mathcal{C}, t \in N_c$\\

$v_{J}$ : target JND\\
\textbf{Output:} $Q$~:Set of selected representations \\

\textbf{Step 1}:\\
\For{each $c \in \mathcal{C}$}{
    t = 2\\
    \While{$t \leq N_{c}$}{
        \If{$\hat{v}_{c, \hat{r}_{t}, \hat{b}_{t}} > v_{\text{max}}$ or $\hat{v}_{c, \hat{r}_{t}, \hat{b}_{t}} - \hat{v}_{c, \hat{r}_{t-1}, \hat{b}_{t-1}} < v_J$}
        {Eliminate $( \hat{r}_{t}, \hat{b}_{t}, c)$ from the ladder\\
        $\hat{N}_{c} = N_{c} - 1$
        }
        $t=t+1$
    }
}

\textbf{Step 2}:\\
$Q$= \{$(\hat{r}_{t}, \hat{b}_{t}, c_{1})$\}, $t \in \hat{N}_{c_{1}}$ \\

\For{each $c \in \{c_{2}, .., c_{M}\}$}{
     \For{each $t \in \hat{N}_{c}$}{
      $(\Tilde{r}_i, \Tilde{b}_i, c_1) \gets \argmin_{i} \mid \hat{b}_{i, c_1} - \hat{b}_{t} \mid s.t.~~ \hat{b}_{t} \geq b_{i, c_1}$\\
      $(\Tilde{r}_j, \Tilde{b}_j, c_1) \gets \argmin_{j} \mid \hat{b}_{j, c_1} - \hat{b}_{t} \mid s.t.~~ \hat{b}_{t} \leq \hat{b}_{i, c_1}$\\
      RD curve $L$ between $(\Tilde{r}_i, \Tilde{b}_i, c_1)$ and $(\Tilde{r}_j, \Tilde{b}_j, c_1)$: $v = \frac{\hat{v}_{c_1, \hat{r}_{j}, \hat{b}_{j}} - \hat{v}_{c_1, \hat{r}_{i}, \hat{b}_{i}}}{\Tilde{b}_j - \Tilde{b}_i}\cdot(b - \Tilde{b}_i) + \hat{v}_{c_1, \hat{r}_{i}, \hat{b}_{i}}$\\
      \If{ ($\hat{v}_{c, \hat{r}_{t}, \hat{b}_{t}}$ is above $L$)}
      {Add $(\hat{r}_{t}, \hat{b}_{t}, c)$ to $Q$. }
    }
}

\vspace{-0.3em}
\caption{Redundant representation elimination.}
\label{algo:jnd_bl_pred}
}
\end{algorithm}
\rf{The pseudo-code of the \textit{redundant representation elimination} method is shown in Algorithm~\ref{algo:jnd_bl_pred}. This algorithm consists of two primary steps as follows:}
% \subsubsection{Step 1}

\rf{\textit{\textbf{Step 1:}}}
In practice, it is often observed that the VMAF scores of different representations are \rf{highly} similar, \rf{leading to} perceptual redundancy in the bitrate ladder. \sam{Consequently, this redundancy implies a wastage of energy during the encoding, storage, and transmission of data, without any improvement in QoE}. To \sam{minimize} this \sam{perceptual redundancy}, \rf{\mcbe} leverages the concept of the JND threshold, which represents the minimum threshold at which the human eye can perceive differences in quality~\cite{lin2015experimental, wang2016mcl, wang2017videoset}. A fixed JND threshold denoted as $v_{J}$ is input from the streaming service provider. If the VMAF difference between two representations \rf{is} lower than $v_{J}$, the higher bitrate representation among them is eliminated. Furthermore, when the predicted VMAF is greater than \sam{the maximum VMAF above which the representation is deemed perceptually lossless} ($v_{\text{max}}$), the corresponding representation is eliminated from the bitrate ladder~\cite{cvfr_ref}. 
% \sam{as shown in Algorithm~\ref{algo:jnd_bl_pred}. 
\sam{This way, \mcbe lowers the overall energy requirement for encoding.}
%Redundant representation elimination is divided into three steps: (i) VMAF prediction of all representations, (ii) JND-based representation elimination, and (iii) 
%\textit{JND-aware elimination}:
% \subsubsection{Step 2}

\rf{\textit{\textbf{Step 2:}}} RD points \sam{(based on the bitrates predicted by the bitrate ladder estimators (\cf Fig.~\ref{fig:mcps_scheme}) and the corresponding predicted VMAF \rf{scores})} of each representation of new-generation codecs are \sam{geometrically} compared to the \sam{predicted} RD curve of the previous generation codec. The representation is eliminated if the point is below the RD curve of the previous generation codec.

%\subsection{Encoding Representations}
%\label{sec:enc}
In the final phase, the encoding process is performed exclusively for the selected bitrate-resolution combinations ($\hat{b}, \hat{r}$) of each codec ($\hat{c}$) for every video segment.

\begin{figure*}[t]
\centering
\begin{subfigure}{0.245\textwidth}
    \centering
   \includegraphics[clip,width=\textwidth]{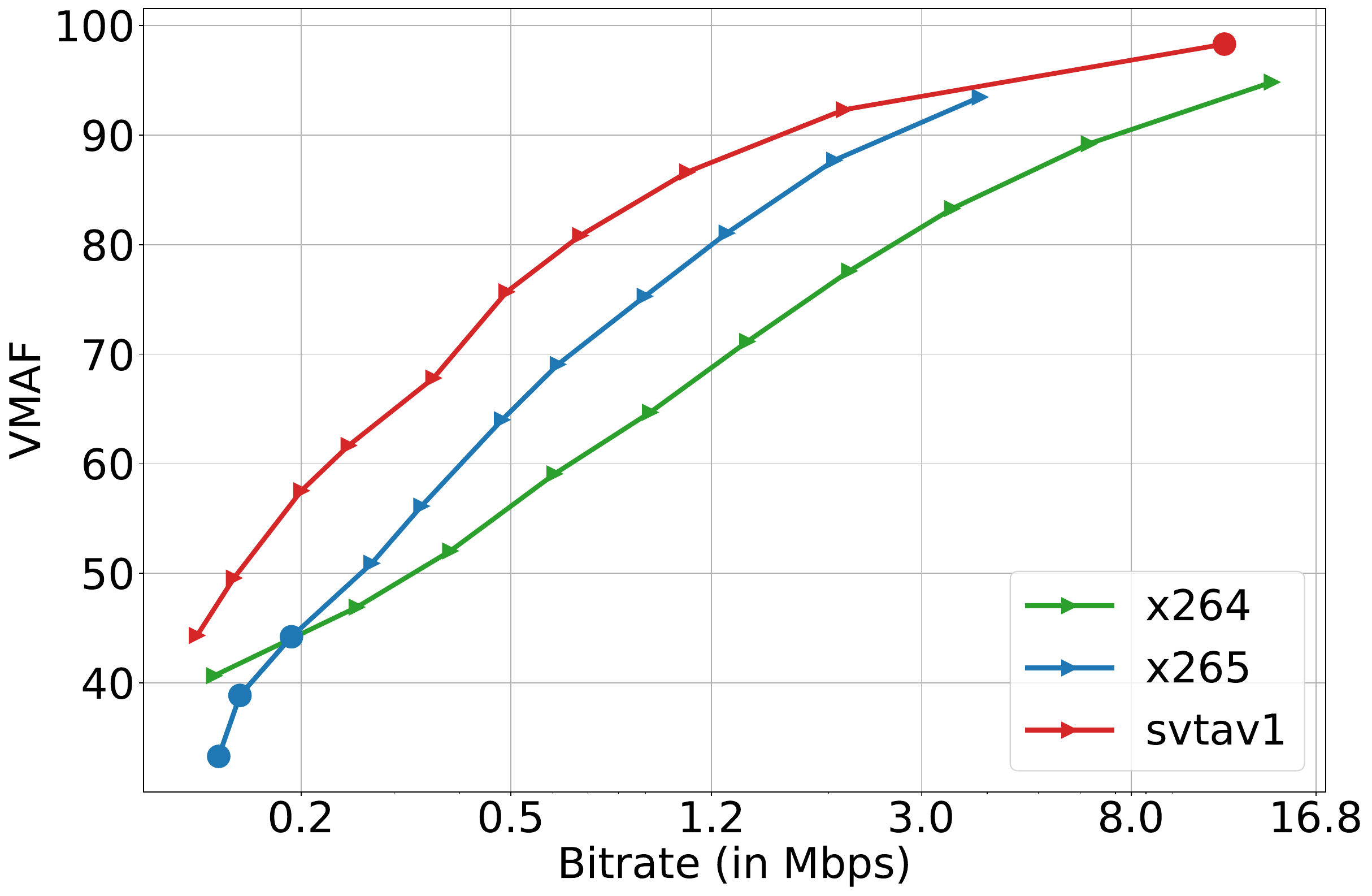}
    \caption{\textit{Bunny\_s000}}
\end{subfigure}
\hfill
\begin{subfigure}{0.245\textwidth}
    \centering
   \includegraphics[clip,width=\textwidth]{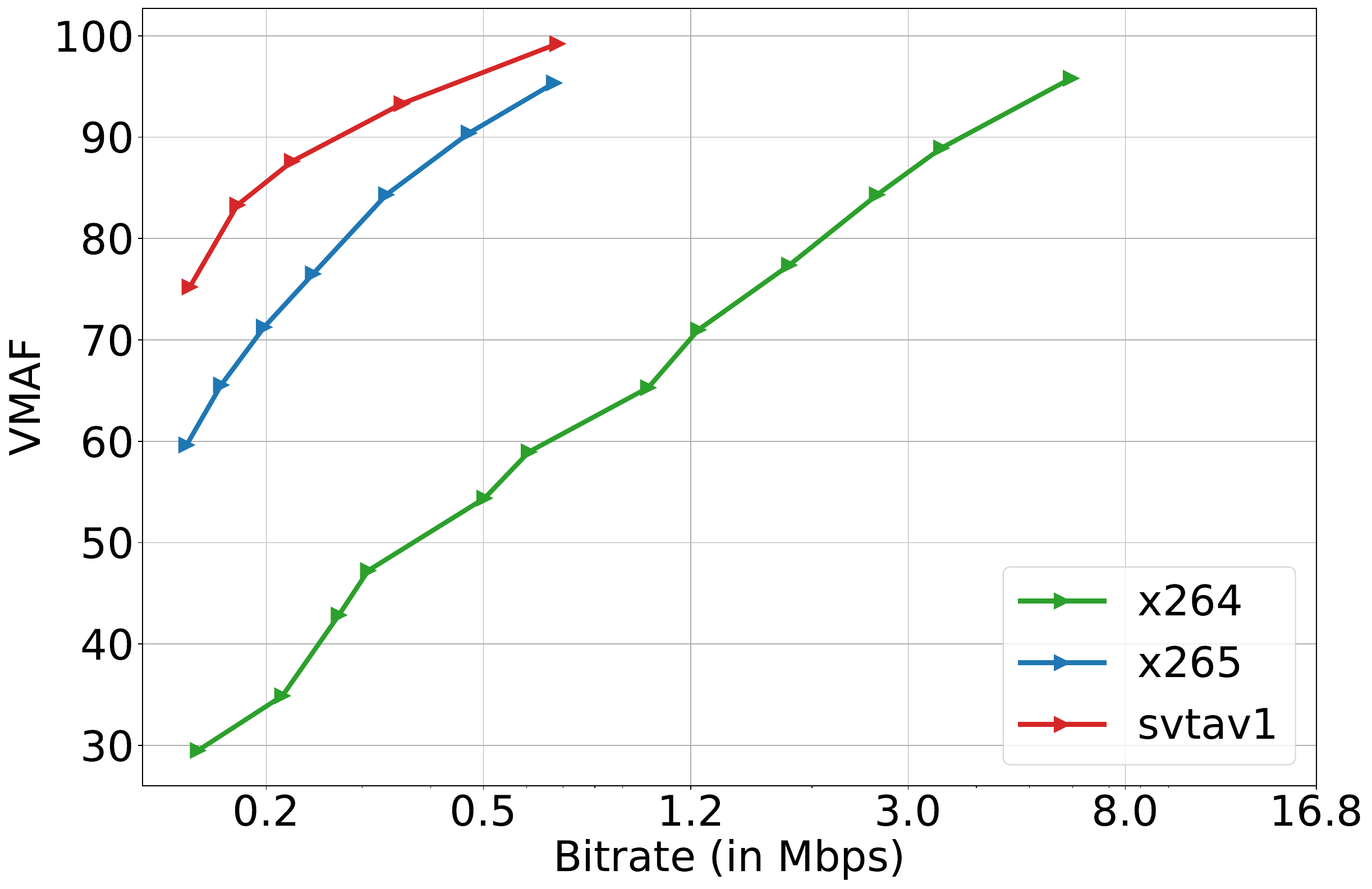}
    \caption{\textit{Characters\_s000}}
\end{subfigure}
\hfill
\begin{subfigure}{0.245\textwidth}
    \centering
   \includegraphics[clip,width=\textwidth]{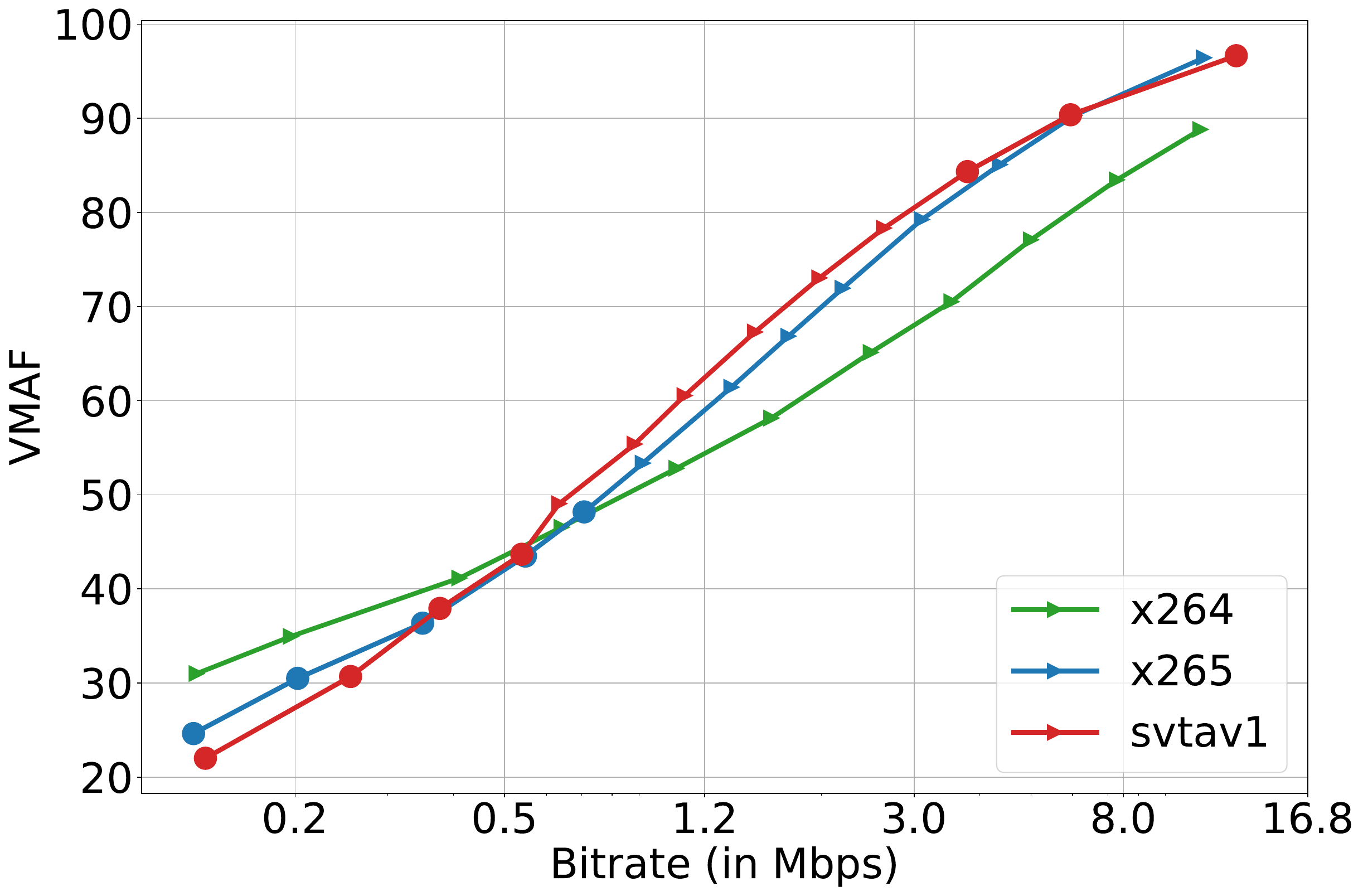}
    \caption{\textit{RushHour\_s000}}
    \label{fig:rushhour_rd1}
\end{subfigure}
\hfill
\begin{subfigure}{0.245\textwidth}
    \centering
   \includegraphics[clip,width=\textwidth]{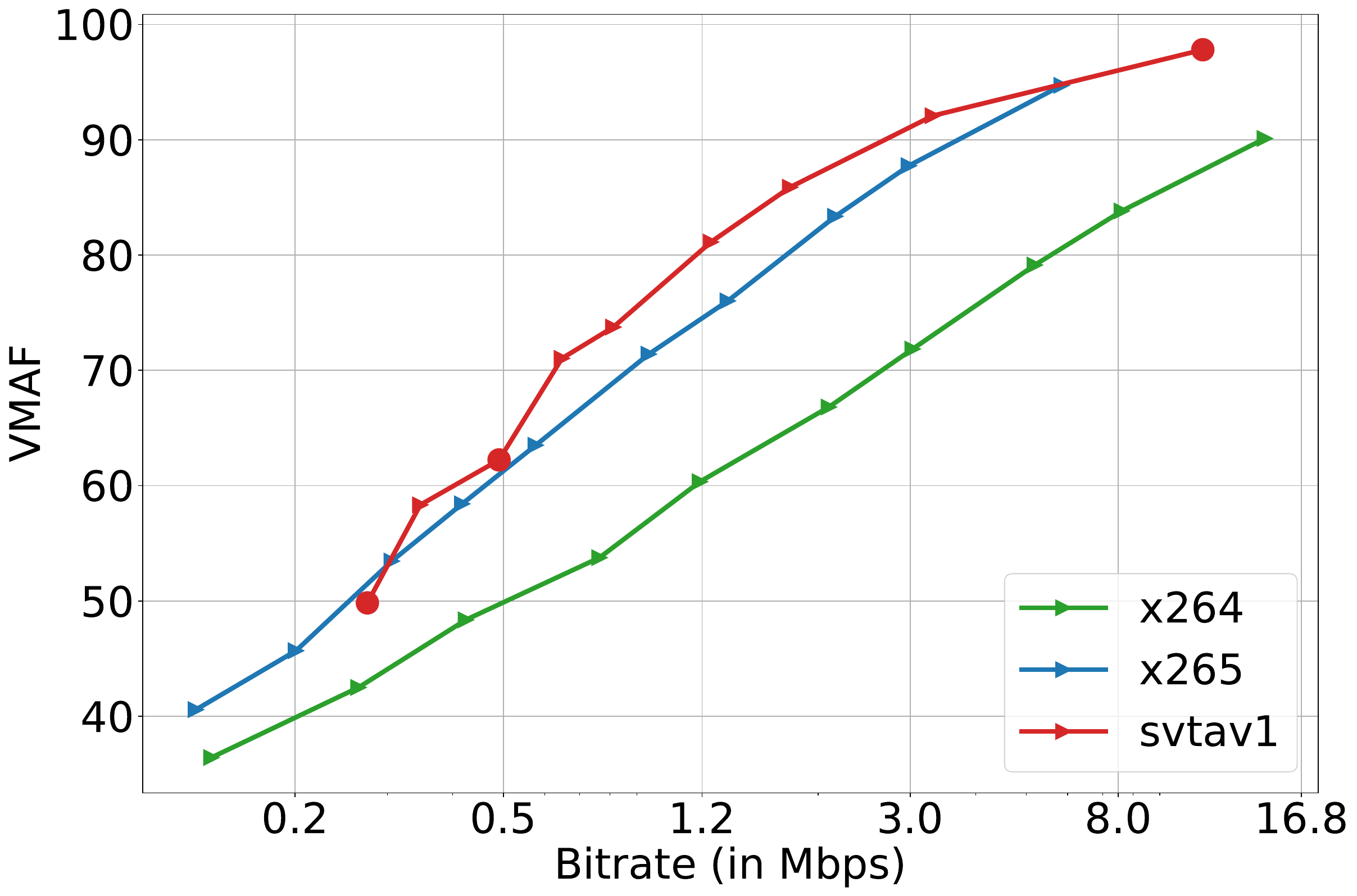}
    \caption{\textit{Wood\_s000}}
    \label{fig:wood_rd1}
\end{subfigure}
\vspace{-0.9em}
\caption{RD curves of representative segments (a) \textit{Bunny\_s000} (\EY=22.40, \h=4.70, \LY=129.21), (b) \textit{Characters\_s000} (\EY=45.42, \h=36.88, \LY=134.56), (c) \textit{RushHour\_s000} (\EY=47.75, \h=19.70, \LY=101.66), and (d) \textit{Wood\_s000} (\EY=124.72, \h=47.03, \LY=119.57) of VCD dataset~\cite{VCD_ref} encoded using \mcbe (x264, x265, svtav1). Here, \etpsplus~\cite{jtps_ref} is considered as the bitrate ladder prediction method, and $v_{J}=6$. Representations marked \sam{using dots} indicate the eliminated representations.}
\vspace{-0.10em}
\label{fig:rd_res_mcps}
\end{figure*}
\begin{table*}[t]
\caption{Average performance results using \mcbe \sam{compared to} HLS, \opte, and \etpsplus bitrate ladders prediction methods for various target encoder combinations.}
\centering
\resizebox{0.79\linewidth}{!}{
\begin{tabular}{l|c||c|c|c|c|c|c|c|c|c}
\specialrule{.12em}{.05em}{.05em}
\specialrule{.12em}{.05em}{.05em}
\multicolumn{2}{c||}{\mcbe configuration}  & \multicolumn{3}{c|}{HLS ladder~\cite{HLS_ladder_ref}} & \multicolumn{3}{c|}{\opte~\cite{opte_ref}} & \multicolumn{3}{c}{\etpsplus~\cite{jtps_ref}}\\
\specialrule{.12em}{.05em}{.05em}
\specialrule{.12em}{.05em}{.05em}
Target encoders & $v_{J}$ & $\Delta E_{\text{enc}}$ & $\Delta S$ & $\Delta E_{\text{sto}}$ & $\Delta E_{\text{enc}}$ & $\Delta S$ & $\Delta E_{\text{sto}}$ & $\Delta E_{\text{enc}}$ & $\Delta S$ & $\Delta E_{\text{sto}}$ \\
\specialrule{.12em}{.05em}{.05em}
\multirow{3}{*}{(x264, x265)}         & 2 & -34.05\% & -43.70\% &  -68.30\% & -36.03\%  & -46.12\% & -70.97\% & -15.82\% & -10.20\% & -19.36\% \\
                                      & 4 & -47.72\% & -60.48\% &  -84.38\% & -49.96\%  & -63.00\% & -86.31\% & -16.07\% & -9.53\% & -18.15\% \\
                                      & 6 & -58.09\% & -69.91\% &  -90.94\% & -59.75\%  & -72.50\% & -92.44\% & -16.18\% & -11.27\% & -21.26\% \\
\hline
\multirow{3}{*}{(x264, svtav1)}       & 2 & -34.50\% & -43.31\% &  -67.87\% & -36.87\%  & -45.73\% & -70.55\% & -12.76\% & -8.79\% & -16.81\% \\
                                      & 4 & -48.19\% & -59.95\% &  -83.96\% & -51.17\%  & -62.82\% & -86.18\% & -12.82\% & -8.27\% & -15.86\% \\
                                      & 6 & -58.61\% & -69.18\% &  -90.50\% & -61.06\%  & -72.60\% & -92.49\% & -12.90\% & -9.38\% & -17.88\% \\
\hline
\multirow{3}{*}{(x264, x265, svtav1)} & 2 & -20.42\% & -53.20\% &  -78.10\% & -18.55\%  & -53.63\% & -78.50\% & -22.56\% & -14.57\% & -27.01\% \\
                                      & 4 & -41.67\% & -69.15\% &  -90.49\% & -39.68\%  & -69.49\% & -90.69\% & -23.74\% & -17.81\% & -32.45\% \\
                                      & 6 & -56.45\% & -77.61\% &  -94.99\% & -54.34\%  & -78.32\% & -95.30\% & -27.80\% & -22.62\% & -40.12\% \\
\specialrule{.12em}{.05em}{.05em}
\specialrule{.12em}{.05em}{.05em}
\end{tabular}}
\vspace{-0.89em}
\label{tab:mcps_res_cons}
\end{table*}

\section{Experimental results }
\label{sec:evaluation}

\subsection{Test Setup}
\label{sec:test_methodology}
In this paper, \rf{400} sequences (80\% of the sequences) from the Video Complexity Dataset~\cite{VCD_ref} are used as the training dataset, and the remaining (\ie 20\%) are used as the test dataset. The sequences are encoded at 30fps with the fastest encoding preset supported by the considered encoders on a dual-processor server with Intel Xeon Gold 5218R (80 cores, frequency at 2.10 GHz) with $\mathcal{C}$=\{x264 v1.1, x265 v3.5, svtav1 v1.6\}. VCA and the encoders specified in $\mathcal{C}$ are run using eight CPU threads with x86 SIMD optimization~\cite{x86_simd_ref}. The resolutions specified in the Apple HLS authoring specifications~\cite{HLS_ladder_ref} are considered in the evaluation, \ie $\mathcal{R}$= \{360p, 432p, 540p, 720p, 1080p, 1440p, 2160p\}. In \rf{all} experiments, the average target JND function ($v_{J}$) is considered as two~\cite{kah_ref}, four, and six~\cite{jnd_streaming} based on current industry practices. \sam{Accordingly, $v_{\text{max}}$ is set as 98, 96, and 94, respectively.} This paper uses the following state-of-the-art encoding bitrate ladder prediction schemes in conjunction with \mcbe:

\begin{enumerate}
    \item Default HLS bitrate ladder~\cite{HLS_ladder_ref} for each codec/encoder.
    \item \opte~\cite{opte_ref}, where optimized resolutions for the set of bitrates in the HLS bitrate ladder are predicted for each encoder.
    \item \etpsplus~\cite{jtps_ref}, where optimized bitrate-resolution pairs are predicted for JND-aware efficient encoding for each encoder.
\end{enumerate}
Note that separate bitrate ladders are generated for each encoder in the state-of-the-art encoding schemes.

\subsection{Latency and Accuracy Analysis}
Spatio-temporal features (\cf Section~\ref{sec:features}) are extracted at a rate of 370\,fps. The overall inference time (including the feature extraction time, VMAF prediction time, and inference time) for a 4\,s video segment of 2160p resolution is 0.37\,s. Hence, the additional latency introduced by \mcbe is negligible. The average mean absolute error (MAE) of VMAF prediction for all resolutions is observed to be 2.42, which is acceptable for live-streaming applications.

\subsection{Storage Consumption Analysis}
Fig.~\ref{fig:rd_res_mcps} shows the rate-distortion (RD) curves of selected video sequences (segments) encoded using \etpsplus bitrate ladder prediction method for x264, x265, and svtav1. It is observed that there are bitrate regions where the new-generation codecs (\rf{\ie} HEVC and AV1) have lower RD performance compared to AVC. \mcbe eliminates the representations of new-generation codecs \rf{when} their \sam{predicted} VMAF is lower than the RD curve of \sam{the} AVC encoding. In Fig.~\ref{fig:rd_res_mcps}, \sam{dot} marks indicate the eliminated representations. Furthermore, it is also observed that \mcbe removed the perceptual redundancy between multiple codec representations based on the JND threshold of six VMAF points. Table~\ref{tab:mcps_res_cons} shows the storage reduction ($\Delta S$) using \mcbe in conjunction with the HLS bitrate ladder, \opte, and \etpsplus. As $v_{J}$ increases, more representations are eliminated, which reduces the storage needed. HLS bitrate ladder and \opte representations have high perceptual redundancy compared to \etpsplus, as \etpsplus representations are predicted with a perceptual gap of one JND~\cite{jtps_ref}. Hence, storage reduction is significantly high with HLS bitrate ladder and \opte, compared to \etpsplus.

\subsection{Energy Consumption Analysis}
\vspace{-0.2em}
This section evaluates energy consumption using \mcbe in terms of \textit{(i)} encoding~($\Delta E_{\text{enc}}$), and \textit{(ii)} storage~($\Delta E_{\text{sto}}$). 
The \textit{CodeCarbon} tool~\cite{codecarbon_ref} is used to calculate the encoding energy. The storage energy is modeled inspired by~\cite{bianco2016energy} as $E_{\text{sto}} =~S_d \cdot P_{\text{b}} \cdot T_{\text{s}}$, where $S_d$ is the video data size (in bits), $P_{\text{b}}$ is power consumption per bit (in \si{\watt / \bit}), and $T_{\text{s}}$ is the time %\footnote{measured using the \texttt{dd} command —a tool to measure disk write speed.}
taken for data to be stored (in hours). %Furthermore, \rf{the} transmission energy is modeled as $E_{tra} = I_{net}  \cdot  (\frac{SIZ_d}{TH_{net}}) \cdot D$, where $I_{net}$ is the energy intensity of the network (in \si{\watt \hour / \giga\byte}), $ TH_{net}$ is the network throughput (in \si{ \giga\byte / \hour}), and $D$ is the transmission rate of data (in \si{ \giga\byte / \hour})~\cite{carbontrust2021}.

Table~\ref{tab:mcps_res_cons} illustrates the average energy reduction achieved in encoding and storage using \mcbe \rf{compared} to the alternative schemes. Negative values in the table indicate the extent of \rf{the} reduction in energy consumption. Compared to the state-of-the-art, the results show significant encoding energy reduction $\Delta E_{\text{enc}}$ for \mcbe. For instance, in a streaming session with devices supporting AVC, HEVC, and AV1 decoding, and considering a JND of six VMAF points, \mcbe achieves energy reductions of up to 56.45\%, 54.34\%, and 27.80\% when compared to HLS bitrate ladder encoding, \opte, and \etpsplus, respectively. This substantial reduction in encoding energy is primarily because \mcbe eliminates the need to encode segments for all representations in all x264, x265, and svtav1 bitrate ladders as explained previously in Section~\ref{sec:rep_elim}. Instead, \mcbe selects the representation with the lowest energy requirement (\ie AVC representation) for encoding when representations in different codecs have the same VMAF value. Consequently, \mcbe predominantly includes all x264 representations and only the higher VMAF representations from the other codecs, leading to a reduced amount of data to encode $\Delta E_{\text{enc}}$ and store $\Delta E_{\text{sto}}$, thus consuming less storage energy. For example, compared to the HLS ladder, \mcbe reduces the data to store by 77.61\%, resulting in a remarkable 94.99\% less energy consumed for storage during a streaming session with devices supporting AVC, HEVC and AV1 decoding and a JND of six VMAF points. %\rf{Energy transmission is similarly affected by the volume of the data}, resulting in $\Delta E_{tra}$ being consistent with $\Delta S$ in all our experiments.

\section{Conclusions}
\label{sec:conclusion_future_dir}
This paper proposed \mcbe, an online energy-efficient multi-codec JND-aware bitrate ladder estimation scheme for adaptive streaming applications. \mcbe includes an algorithm to determine an optimized multi-codec encoding bitrate ladder, where redundant representations of new-generation video codecs are eliminated. Furthermore, perceptual redundancy within the representations of each codec is minimized by eliminating representations based on \rf{the} JND threshold. \mcbe can be used in conjunction with any bitrate ladder estimation scheme. \mcbe on average, yields encoding, storage, and transmission energy savings of 56.45\%, 77.61\%, and 94.99\%, respectively, compared to the state-of-the-art HLS bitrate ladder encoding, for a streaming session with devices supporting AVC, HEVC, and AV1 decoding, considering a JND of six VMAF points.
\balance
%\section*{Acknowledgement}
%We gratefully acknowledge
%\emph{Austrian Research Promotion Agency (FFG)}, grant agreement FO999897846 (GAIA), and the Christian Doppler Research Association. Christian Doppler Laboratory ATHENA: {https://athena.itec.aau.at/}.
\iffalse
In this work, we only covered a part of the entire end-to-end
streaming chain (encoding, storage and transmission) since there exists no generic model for the user-side
streaming power usage. The user device energy consumption depends on various browsers, operating systems, codecs,  hardware, or software decoding. 
\fi
\newpage
\bibliography{references.bib}{}
\bibliographystyle{IEEEtran}
\balance
\end{document}